\documentclass[11pt,a4paper]{article}
\usepackage{authblk}
\usepackage[hyperref]{acl2021}

\usepackage{times}
\usepackage{latexsym}

\usepackage{microtype}

\aclfinalcopy % Uncomment this line for the final submission
\usepackage{comment}  
\usepackage{soul}
\usepackage[most]{tcolorbox}
\usepackage{enumitem}
\usepackage{latexsym}
\usepackage{numprint}
\usepackage{todonotes}
\usepackage{booktabs}
\usepackage{multirow}
\usepackage{amsmath}
\usepackage{listings}
\lstdefinelanguage
   [x64]{Assembler}     % add a "x64" dialect of Assembler
   [x86masm]{Assembler} % based on the "x86masm" dialect
   {morekeywords={CDQE,CQO,CMPSQ,CMPXCHG16B,JRCXZ,LODSQ,MOVSXD, %
                  POPFQ,PUSHFQ,SCASQ,STOSQ,IRETQ,RDTSCP,SWAPGS, %
                  rax,rdx,rcx,rbx,rsi,rdi,rsp,rbp, %https://it.overleaf.com/project/5e7bd8436be44f000172b14f
                  r8,r8d,r8w,r8b,r9,r9d,r9w,r9b, %
                  r10,r10d,r10w,r10b,r11,r11d,r11w,r11b, %
                  r12,r12d,r12w,r12b,r13,r13d,r13w,r13b, %
                  r14,r14d,r14w,r14b,r15,r15d,r15w,r15b,}} % etc.

\usepackage{xcolor}
\usepackage[hashEnumerators,smartEllipses]{markdown}

\definecolor{codegreen}{rgb}{0,0.6,0}
\definecolor{codegray}{rgb}{0.5,0.5,0.5}
\definecolor{codepurple}{rgb}{0.58,0,0.82}
\definecolor{backcolour}{rgb}{0.95,0.95,0.92}

\lstdefinestyle{mystyle}{
    backgroundcolor=\color{backcolour},   
    commentstyle=\color{codegreen},
    keywordstyle=\color{blue},
    numberstyle=\tiny\color{codegray},
    stringstyle=\color{codepurple},
    basicstyle=\ttfamily\footnotesize,
    breakatwhitespace=false,         
    breaklines=true,                 
    captionpos=b,                    
    keepspaces=true,                 
    numbers=left,                    
    numbersep=5pt,                  
    showspaces=false,                
    showstringspaces=false,
    showtabs=false,                  
    tabsize=2,
    columns=fullflexible,
}

\newcommand{\datasetname}[1]{\emph{Shellcode\_IA32}}
\usepackage{cancel}

\title{\datasetname{}: A Dataset for Automatic Shellcode Generation}

\author[1]{\textbf{Pietro Liguori}}
\author[2]{\textbf{Erfan Al-Hossami}}
\author[1]{\textbf{Domenico Cotroneo}}
\author[1]{\textbf{Roberto Natella}}
\author[2]{\\ \textbf{Bojan Cukic}}
\author[2]{\textbf{Samira Shaikh}}

\affil[1]{University of Naples Federico II \\
Naples, Italy
}

\affil[2]{University of North Carolina at Charlotte\\
Charlotte, NC, USA}

\affil[ ]{\texttt{\{pietro.liguori, cotroneo, roberto.natella\}@unina.it} 
}

\affil[ ]{\texttt{\{ealhossa, bcukic, samirashaikh\}@uncc.edu} 
}

\date{}

\begin{document}

\onecolumn 

Please refer to the extended journal version of this work. The extended paper provides additional information on the \datasetname{} dataset, and an extensive experimental analysis.

\vspace{2cm}

\begin{tcolorbox}[colback=red!5!white,colframe=red!75!black]

\noindent
{\Large \textit{Can we generate shellcodes via natural language? An empirical study.}}

\vspace{5pt}
\noindent
Pietro Liguori, Erfan Al-Hossami, Domenico Cotroneo, Roberto Natella, Bojan Cukic, and Samira~Shaikh

\vspace{5pt}
\noindent
Automated Software Engineering, Volume 29, Article no. 30, March 2022

\vspace{5pt}
\noindent
DOI: \href{https://doi.org/10.1007/s10515-022-00331-3}{10.1007/s10515-022-00331-3}

\vspace{5pt}
\noindent
arXiv: \href{https://arxiv.org/abs/2202.03755}{2202.03755}

\end{tcolorbox}

\newpage
\maketitle

\begin{abstract}
We take the first step to address the task of automatically generating shellcodes, i.e., small pieces of code used as a payload in the exploitation of a software vulnerability, starting from natural language comments. 
We assemble and release a novel dataset (\textit{Shellcode\_IA32}), consisting of challenging but common assembly instructions with their natural language descriptions. We experiment with standard methods in neural machine translation (NMT) to establish baseline performance levels on this task.
\end{abstract}

\section{Introduction and Related Work}
\label{sec:introduction}
A growing body of research has dealt with automated code generation: given a natural language description, a code comment or intent, the task is to generate a piece of code in a programming language  \cite{DBLP:journals/corr/YinN17,DBLP:journals/corr/LingGHKSWB16}. 
The task of generating programming code snippets, also referred to as semantic parsing \cite{yin2019reranking,xu-etal-2020-incorporating}, has been previously addressed to generate executable snippets in domain-specific languages~\cite{guu2017language,long2016simpler}, and several programming languages, including Python \cite{DBLP:journals/corr/YinN17} and Java \cite{DBLP:journals/corr/LingGHKSWB16}.

% %  Erfan: Adds here reviewer paper descriptions.
% SCONE~\cite{guu2017language,long2016simpler} is a semantic parsing dataset, focusing on generating executable actions from natural language instructions in 3 different domains. Our similarity with it is that our natural language instructions are very specific within the context and results in very small actions that add up to completing a bigger goal. Our work is different in that we are generating program snippets in the assembly that are used for exploits. Their focus is understanding human instructions even when they contain pronouns or incomplete sentences in different domains, while our work focuses on the sole domain of generating assembly snippets.

% Cai \textit{et al.}~\citeyearpar{cai2017making} explore using recursion in a low-level programming language for synthesizing programs on four tasks, grade-school addition, topological sort, bubble sort, and quicksort.

We consider the task of generating \emph{shellcodes}, i.e., small pieces of code used as a payload to exploit software vulnerabilities. Shellcoding, in its most literal sense, means writing code that will return a remote shell when executed. It can represent any byte code that will be inserted into an exploit to accomplish the desired, malicious, task \cite{mason2009english}. An example of a shellcode program in assembly language and the corresponding natural language comments are shown in Listing~\ref{list:shellcode}.

Shellcodes are important because they are the key element of security attacks: they represent code injected into victim software to take control of a machine, to escalate privileges, and to use the machine for malicious purposes such as DDoS attacks, data theft, and running malware \cite{arce2004shellcode}. Well-intentioned actors (security practitioners and product vendors) also develop shellcodes to run non-harmful \emph{proof-of-concept} attacks, to show how security weaknesses can be exploited to identify vulnerabilities and patch systems. Thus, shellcode generation using (semi-) automated techniques has become a popular and very active research topic \cite{bao2017your}. 
%https://searchsecurity.techtarget.com/definition/proof-of-concept-PoC-exploit
However, writing shellcodes is technically challenging since they are typically written in assembly language (c.f. Listing~\ref{list:shellcode}). The most sophisticated shellcodes can reach hundreds of assembly lines of code.

\begin{figure}[t!]
\footnotesize
\begin{minipage}{\linewidth}
\begin{lstlisting}[caption={x86 assembly code used to spawn \texttt{/bin/sh} shell on Linux OS. Lines 4-5, 6-7-8, 9-10, 11-12 are multi-line snippets generated by four different intents.},label={list:shellcode}]
global _start;  Declare global _start.
section .text;  Declare the text section.
_start:;        Define the _start label.
xor eax, eax;   Zero out the eax                    register 
push eax;       and push its contents               on the stack.
push 0x68732f2f;Move /bin//sh
push 0x6e69622f;into the ebx register.
mov ebx, esp
push eax;       Push the contents of eax            onto the stack 
mov edx, esp;   and point edx to the                stack register.
push ebx;       Push the contents of ebx            onto the stack
mov ecx, esp;   and point ecx to the                stack register.
mov al, 11;     Put the system call 11              into the al register.
int 0x80;       Make the kernel call.
\end{lstlisting}
%\vspace{-0.6cm}
\end{minipage}
\end{figure}

The task of the shellcode generation has been addressed by several works and tools. Bao \textit{et al.} (\citeyear{bao2017your}) designed ShellSwap, a system that can modify an observed exploit and replace the original shellcode with an arbitrary replacement shellcode. The system uses symbolic tracing, with a combination of shellcode layout remediation and path kneading to achieve shellcode transplants. 
\textit{Pwntools} \cite{pwntools} is a CTF framework and exploit development library written in Python. It is designed for rapid prototyping and development and intended to make exploit writing as simple as possible.

%Our work is distinct from previous work in code generation from natural language, in that we focus on generating snippets for the domain of cybersecurity and in particular assembly-based exploit generation. 
Differently from previous work in the security literature, we approach this problem as a machine translation (NMT) task. We apply neural machine translation \cite{goodfellow2016deep}, which unlike the traditional phrase-based translation system consisting of many small sub-components tuned separately, attempts to build and train a single, large neural network that reads a sentence and outputs a correct translation \cite{bahdanau2014neural}. NMT has emerged as a promising machine translation approach, showing superior performance on public benchmarks \cite{bojar2016findings}, and it is widely recognized as the premier method for the translation of different languages \cite{wu2016google}. 
NMT has also been used to perform complex tasks on the UNIX operating system shell \cite{lin2017program} (e.g. file manipulation and search), by stating goals in English \cite{lin-etal-2018-nl2bash}, to automatically generate commit messages \cite{liu2018neural}, etc. However, the NMT techniques have not heretofore been adopted to automatically generate software exploits from natural language comments.

Since NMT is a data-driven approach to code generation, we need a dataset of intents in natural language, and their corresponding translation (in our context, in assembly language) for shellcode generation. 
In this preliminary work, we address the lack of such a dataset by presenting \datasetname{}, a dataset containing $3,200$ lines of assembly code extracted from real shellcodes and described in the English language. Moreover, we present experiments on our dataset using a baseline technique, in order to establish performance levels for evaluating shellcode generation techniques. 

%In the following, Section~\ref{sec:dataset} describes \datasetname{}; Section~\ref{sec:experiments} shows preliminary experiments performed on \datasetname{} dataset; Section~\ref{sec:ethics} discusses the ethical considerations; Section~\ref{sec:conclusion} concludes the paper.

\section{Dataset}
\label{sec:dataset}
We compiled a dataset, \textbf{\datasetname{}}, specific to our task. This dataset consists of \numprint{3,200} examples of instructions in assembly language for \textit{IA-32} (the 32-bit version of the x86 Intel Architecture) from publicly-available security exploits. We collected assembly programs used to generate shellcode from \textit{shell-storm} \cite{shellstorm} and from \textit{Exploit Database} \cite{exploitdb}, in the period between 2000 and 2020.

\begin{figure}[h!]
    \centering
    \includegraphics[width=\columnwidth]{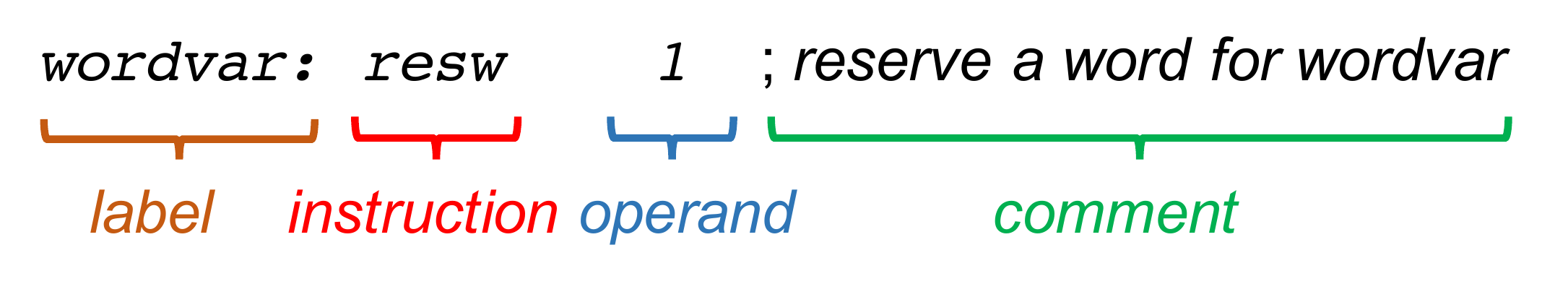}
    %\vspace{-0.8cm}
    \caption{Layout of NASM source line}
    \label{fig:assembly_instruction}
    %\vspace{-0.25cm}
\end{figure}

Our focus is on Linux, the most common OS for security-critical network services. Accordingly, we added assembly instructions written with \textit{Netwide Assembler} (NASM) for Linux \cite{duntemann2000assembly}.
NASM is line-based. Figure~\ref{fig:assembly_instruction} shows a simple example of a NASM source line. Every source line contains a combination of four fields: an optional \textit{label} used to represent either an identifier or a constant, a \textit{mnemonic} or \textit{instruction}, which identifies the purpose of the statement and followed by zero or more \textit{operands} specifying the data to be manipulated, and an optional \textit{comment}, i.e., text ignored by the compiler. A mnemonic is not required if a line contains only a label or a comment. 
 
Each line of \datasetname{} dataset represents a snippet~\textendash~intent pair. The \textbf{snippet} is a line or a combination of multiple lines of assembly code, built by following the NASM syntax. The \textbf{intent} is a comment in the English language (c.f. Listing~\ref{list:shellcode}).

To take into account the variability of descriptions in natural language, multiple authors described independently different samples of the dataset in the English language. Where available, we used as natural language descriptions the comments written by developers of the collected programs. %This allowed us to take into account the variability of descriptions in natural language.
We enriched the dataset by adding examples of assembly programs for the IA-32 architecture from popular tutorials and books \cite{duntemann2011assembly, kusswurm2014modern, tutorialspoint} to understand how different authors and assembly experts describe the code and, thus, how to deal with the ambiguity of natural language in this specific context. 
Our dataset consists of $\sim10\%$ of instructions collected from books and guidelines and the rest from real shellcodes. %There is no qualitative difference between both sets.

\noindent
\textbf{Multi-line Snippets:} To automatically generate shellcodes, we need to look beyond a one-to-one mapping between a line of code and its comment/intent. For example, a common operation in shellcodes is to save the ASCII \textit{``/bin/sh"} into a register. This operation requires three distinct assembly instructions: push the hexadecimal values of the words \textit{``/bin"} and \textit{``//sh"} onto the stack register before moving the contents of the stack register into the destination register (lines 6-8 in Listing~\ref{list:shellcode}). It would be meaningless to consider these three instructions as separate. To address such situations, we include  $510$ lines ($\sim16\%$ of the dataset) of intents that generate multiple lines of shellcodes (separated by the newline character \textbackslash{n}). Table~\ref{tab:multiple_snippets} shows two further examples of multi-line snippets with their natural language intent. 
\begin{table}[t]
\centering
\footnotesize
\begin{tabular}{@{}l@{}}
\toprule
\begin{tabular}[c]{@{}l@{}}\textbf{Intent:} \textit{jump short to the decode label if the contents of the} \\ \textit{\texttt{al} register is not equal to the contents of the \texttt{cl} register} \\ \textit{else jump to the shellcode label} \\ \textbf{Multi-line Snippets:} \texttt{cmp al, cl} \textbackslash{}n \\ \texttt{jne short decode} \textbackslash{}n  \texttt{jmp shellcode}\end{tabular} \\ \midrule
\begin{tabular}[c]{@{}l@{}}\textbf{Intent:} \textit{jump to the label recv\textunderscore http\textunderscore request} \\ \textit{if the contents of the \texttt{eax} register is not zero else subtract} \\ \textit{the value \texttt{0x6} from the contents of the \texttt{ecx} register}\\ \textbf{Multi-line Snippets:}
\texttt{test eax, eax} \textbackslash{}n \\ \texttt{jnz recv\textunderscore http\textunderscore request} \textbackslash{}n   \texttt{sub ecx, 0x6}\end{tabular} \\ \bottomrule
\end{tabular}
\caption{Examples of multi-line snippets}
%\vspace{-0.35cm}
\label{tab:multiple_snippets}
\end{table}

\begin{figure}[t]
    \centering
    \includegraphics[width=\linewidth]{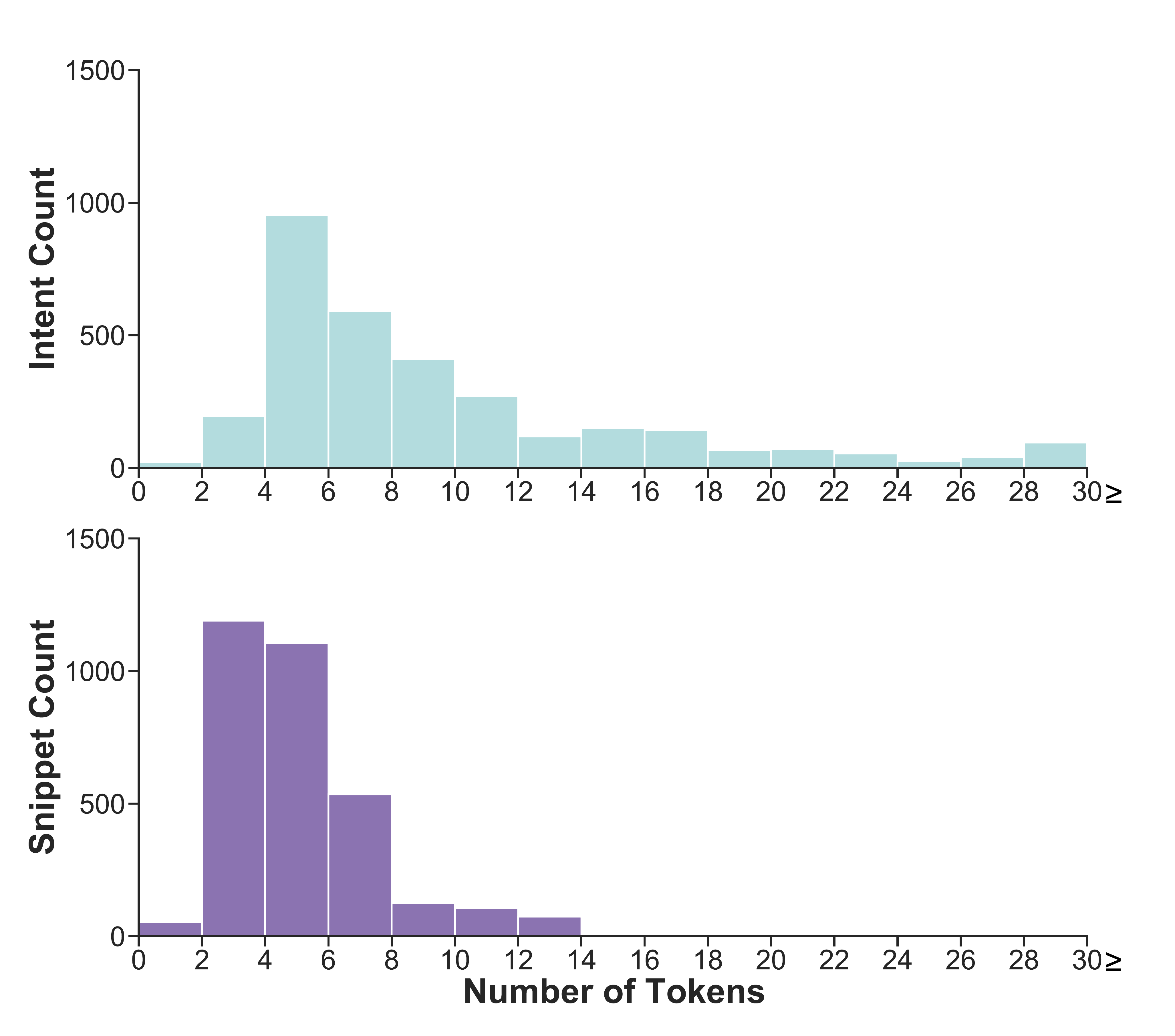}
    \caption{Histogram of the \datasetname{} dataset showcasing the distribution of token counts across intents and snippets.}
    \label{fig:data_hist}
\end{figure}

\noindent
\textbf{Statistics:} Table \ref{tab:dataset_statistics} presents the descriptive statistics of the \datasetname{} dataset.
The dataset contains $52$ distinct assembly instructions (excluding function, section, and label declaration).  
The two most frequent assembly instructions are \texttt{mov} ($\sim30$\% frequency), used to move data into/from registers/memory or to invoke a system call, and \texttt{push} ($\sim22$\% frequency), which is used to push a value onto the stack. The next most frequent instructions are the \texttt{cmp} ($\sim 7\%$ frequency), \texttt{xor} and \texttt{jmp} instructions ($\sim 4\%$ frequency). 
%974 mov; 779 push;228 cmp;147 xor;146 jmp
%13337 assembly tokens - 1401 unique
%29282 NL tokens - 1639 unique
%974 low freq assembly
%1074 low freq nl
The \textit{low-frequency words} (i.e., the words that appear only once or twice in the dataset) contribute to the $3.6\%$ and $7.3\%$ of the natural language and the assembly language, resp. 
\figurename{}~\ref{fig:data_hist} shows the distribution of the number of tokens across the intents and snippets in the dataset.
We publicly share our entire \datasetname{} dataset on a GitHub repository.\footnote{The dataset can be found here: \url{https://github.com/dessertlab/Shellcode_IA32}}

\noindent
\textbf{Size of our dataset:} Our dataset contains $3,200$ instances, which may seem relatively small compared to training data available for most common NLP tasks. We note, however, that our dataset is comparable in size to the CoNaLa annotated dataset ($2,379$ training and $500$ test examples), which is one of the standard datasets in code generation  (for English-Python code generation) \cite{yin2018mining}. Further, \datasetname{} contains a higher percentage of multi-line snippets ($\sim16\%$ vs. $\sim4\%$). We also note here that existing code generation datasets do contain a larger, potentially noisy, subset of training examples (ranging in several thousand) obtained by mining the web. For example, the CoNaLa \textit{mined} (as opposed to the CoNaLa \textit{annotated}) dataset contains 598,237 training examples mined directly from  Stack Overflow~\cite{yin2018mining}. In our case, although shellcodes are written in assembly language, it is not feasible to simply mine examples of natural language\textendash assembly from the web: not all assembly programs are shellcodes.
Thus, our \datasetname{} dataset, which contains $\sim20$ years of shellcodes from a variety of sources is the largest collection of shellcodes in assembly available to date.

\begin{comment}

\begin{table*}[t]
\footnotesize
\centering
\begin{tabular}{
 >{\centering\arraybackslash}p{1.5cm}
 >{\centering\arraybackslash}p{1.5cm}
 >{\centering\arraybackslash}p{1cm}
 >{\centering\arraybackslash}p{2cm}}
\toprule
\textbf{Language} & \textbf{Unique statements} & \textbf{Unique tokens} & \textbf{Avg. tokens per statement} & \textbf{Min tokens per statement} &\textbf{Max tokens per statement}\\
\midrule
\textit{Natural Language} & 3,184 & 1639 & 9.15\\
\midrule
\textit{Assembly Language} & 2,248 & 1401 & 4.17\\
\bottomrule
\end{tabular}
\caption{\datasetname{} statistics}
\label{tab:dataset_statistics}
\end{table*}
\end{comment}

\begin{table}[t]
\footnotesize
\centering
\begin{tabular}{
 >{\centering\arraybackslash}m{3cm} |
 >{\centering\arraybackslash}m{1.5cm}
 >{\centering\arraybackslash}m{1.5cm}}
\toprule
\textbf{Statistics} & \textbf{Natural Language} & \textbf{Assembly Language}\\
\midrule
\textit{Unique Statements} & 3,184 & 2,248\\
\midrule
\textit{Unique Tokens} & 1,498 & 1,244\\
\midrule
\textit{Avg. tokens per statement} & 9.22 & 4.38\\
\midrule
\textit{Min tokens per statement} & 1 & 2\\
\midrule
\textit{Max tokens per statement} & 46 & 30\\
\bottomrule
\end{tabular}
\caption{\datasetname{} statistics}
\label{tab:dataset_statistics}
%\vspace{-0.35cm}
\end{table}

\section{Preliminary Evaluation}
\label{sec:experiments}
\begin{table*}[ht!]
\small
\centering
\begin{tabular}{
>{\centering\arraybackslash}m{1.5cm}
>{\centering\arraybackslash}m{1.5cm}
>{\centering\arraybackslash}m{1.5cm}
>{\centering\arraybackslash}m{1.5cm}
>{\centering\arraybackslash}m{1.5cm}
>{\centering\arraybackslash}m{1.5cm}
>{\centering\arraybackslash}m{1.5cm}}
\toprule
\textbf{Number Layers} & \textbf{Layer Dimension}  & \textbf{BLEU-1 (\%)} & \textbf{BLEU-2 (\%)} & \textbf{BLEU-3 (\%)} & \textbf{BLEU-4 (\%)} & \textbf{ACC (\%)} \\ \midrule

\multicolumn{1}{c|}{\multirow{4}{*}{1}} & 64 & 75.75 & 69.76 & 65.14 & 60.8 & 34.69\\
\multicolumn{1}{l|}{} & 128 & 80.80 & 76.29 & 73.10 & 69.69 & 42.5\\
\multicolumn{1}{l|}{} & 256 & 75.50 & 70.50 & 66.65 & 62.86 & 43.75\\
\multicolumn{1}{l|}{} & 512 & \textbf{83.55} & \textbf{80.08} & \textbf{78.06} & \textbf{76.12} & \textbf{51.25}\\ \midrule

\multicolumn{1}{c|}{\multirow{4}{*}{2}} & 64 & 63.25 & 53.24 & 46.12 & 39.46 & 15.62\\
\multicolumn{1}{l|}{} & 128 & 71.79 & 64.24 & 58.25 & 51.65 & 26.25\\
\multicolumn{1}{l|}{} & 256 & 75.13 & 68.63 & 63.94 & 58.93 & 25.62\\
\multicolumn{1}{l|}{} & 512 & \textbf{80.22} & \textbf{75.00} & \textbf{71.11} & \textbf{67.24} & \textbf{43.44}\\ \midrule

\multicolumn{1}{c|}{\multirow{4}{*}{3}} & 64 & 61.98 & 50.68 & 43.02 & 36.15 & 9.38\\
\multicolumn{1}{l|}{} & 128 & 69.75 & 61.08 & 55.09 & 49.18 & 19.06\\
\multicolumn{1}{l|}{} & 256 & \textbf{76.93} & \textbf{71.32} & \textbf{67.41} & \textbf{63.50} & \textbf{31.87}\\
\multicolumn{1}{l|}{} & 512 & 74.99 & 68.58 & 64.23 & 60.36 & 29.38\\ \midrule

\multicolumn{1}{c|}{\multirow{4}{*}{4}} & 64 & 61.41 & 50.68 & 43.58 & 37.33 & 10.00\\
\multicolumn{1}{l|}{} & 128 & 63.26 & 51.98 & 44.62 & 37.57 & 10.94\\
\multicolumn{1}{l|}{} & 256 & 66.94 & 57.85 & 51.97 & 46.87 & 15.31\\
\multicolumn{1}{l|}{} & 512 & \textbf{70.51} & \textbf{62.44} & \textbf{56.27} & \textbf{50.15} & \textbf{18.75}\\ \bottomrule
\end{tabular}
\caption{Performance results obtained by varying the model hyper-parameters. The best performances for each number of layers are in bold.}
\label{tab:ablation_study}
%\vspace{-0.35cm}
\end{table*}

We performed a set of preliminary experiments with our dataset, in order to assess the applicability of NMT in the context of shellcode generation and to establish baseline performance levels for evaluating techniques for future research. 
Similar to the encoder-decoder architecture with attention \cite{bahdanau2014neural}, we use a bi-directional LSTM as the encoder to transform an embedded intent sequence $E = |e_1,...,e_{T_S}|$ into a vector $c$ of hidden states with equal length. We implement this architecture with Bahdanau-style attention \cite{bahdanau2014neural} using {\fontfamily{qcr}\selectfont xnmt} \cite{neubig-etal-2018-xnmt}. We use an Adam optimizer \cite{kingma2014adam} with $\beta_1=0.9$ and $\beta_2=0.999$. The last step is inference. During inference, the auto regressive inference component uses beam search with a beam size of $5$. The train/dev/test split is train (N = 2560), dev (N = 320), and test (N = 320) using a random $80/10/10$ ratio. The test set includes $44$ multi-line snippets ($13.75\%$ of the test set).

Following prior work in this area \cite{DBLP:journals/corr/LingGHKSWB16,DBLP:journals/corr/YinN17,oda2015learning}, we evaluate the translation performance in terms of averaged token level BLEU scores \cite{papineni2002bleu}.
BLEU uses the modified form of n-grams precision and length difference penalty to evaluate the quality of the output generated by the model compared to the referenced one. BLEU measures translation quality by the accuracy of translating ngrams to n-grams, for values of n usually ranging between $1$ and $4$ \cite{han2016machine,munkova2020evaluation}.
%Since a higher BLEU score does not necessarily reflect higher semantic accuracy \cite{tran2019does}, 
We measure the performance of the evaluation task also in terms of exact match accuracy (ACC), which is the fraction of exactly matching samples between the predicted output and the reference  \cite{DBLP:journals/corr/YinN17}. 
Both metrics range between $0$ and $1$.

During our experiments, we set a basic configuration of the model: $\alpha = 0.001$, layers = $1$, vocabulary size = $4,000$, epochs (with early stopping enforced) = $200$, beam size = $5$, \textit{minimun word frequency} = $1$. Next, we performed experiments by varying the dimensionality of the layers from $64$ to $1024$, and the number of layers from $1$ to $4$ while keeping all other hyper-parameters constant. 
Table \ref{tab:ablation_study} summarizes the results. We notice that increasing the number of layers leads to worse performance, while a layer dimension set between $256$ and $512$ is found to be the best option. 

All experiments were performed on a Linux OS running on a virtual machine with $8$ CPU cores and $8$ GB RAM. The computational times are highly dependent on the model hyper-parameters, and range between few minutes to $\sim105$ minutes, with the average training time equal to $\sim28$ minutes.

 \begin{comment}

\begin{table*}[ht!]
\small
\centering
\begin{tabular}
{>{\centering\arraybackslash}m{1.5cm} | 
>{\centering\arraybackslash}m{1.5cm}
>{\centering\arraybackslash}m{1.5cm}  
>{\centering\arraybackslash}m{1.5cm}
>{\centering\arraybackslash}m{1.5cm}
>{\centering\arraybackslash}m{1.5cm}}

\toprule
\textbf{Model} & 
\textbf{BLEU-1 (\%)} & \textbf{BLEU-2 (\%)} & \textbf{BLEU-3 (\%)} & \textbf{BLEU-4 (\%)} & \textbf{ACC (\%)}  \\
\midrule
\textit{Best hyperparameter model} &  &  & &  & \\
\midrule
\textit{TL with CoNaLa} &  &  &  &  &  \\
\midrule
\textit{TL with Django} & & & & & \\
\bottomrule
\end{tabular}
\caption{Performance across metrics after transfer learning (TL) using CoNaLa and Django datasets}
\label{tab:tl}
\end{table*}
\end{comment}

\section{Qualitative Analysis}
\label{sec:qualitative}

\begin{table*}[ht!]
\footnotesize
\centering
\begin{tabular}{
 >{\centering\arraybackslash}m{6.5cm}|
 >{\centering\arraybackslash}m{4cm}|
 >{\centering\arraybackslash}m{4cm}}

\toprule
\textbf{Natural Language} & \textbf{Ground Truth} & \textbf{Model Output}\\
\midrule
\textit{Put ASCII /bin/sh into eax} & \texttt{push 0x68732f2f \textbackslash n push 0x6e69622f \textbackslash n mov eax , esp} & \texttt{push 0x68732f2f \textbackslash n push 0x6e69622f \textbackslash n mov eax, esp} \\
\midrule
%\textit{If the contents of the eax register is not zero then jump to the label l1 else call the function myfunct} & \texttt{test eax , eax \textbackslash n jnz l1 \textbackslash n call myfunct} & \texttt{test eax, eax \textbackslash n jnz l1 \textbackslash n call myfunct} \\
\textit{Place address buff into esi} & \texttt{mov esi, buff} & \texttt{lea esi, [buff]}\\
\midrule
\textit{Perform a bit-wise inversion of edx} &    \texttt{not edx} & \texttt{\textcolor{red}{\textbf{and}} edx\textcolor{red}{\textbf{, 0}}}\\
\midrule
%\textit{Zero out eax and ecx} & \texttt{xor ecx, ecx \textbackslash n mul ecx} & \texttt{xor ecx, ecx} \\
\textit{if the contents of the bl register is greater than or equal to the value 78h then jump to the memory location loc\_402B1D} &	\texttt{cmp bl, 78h  \textbackslash n jge short loc\_402B1D} & \texttt{cmp bl, 78h  \textbackslash n \textcolor{red}{\textbf{jle}} short loc\_402B1D} \\
\bottomrule
\end{tabular}
\caption{Illustrative examples of correct and incorrect output. The prediction errors are \textcolor{red}{\textbf{red/bold}}.}
\label{tab:cases}
\end{table*}

Automated metrics (BLEU and accuracy) provide a somewhat limited window into the efficacy of the models to accomplish our task: the task of automatically generating assembly code from natural language intents. We conducted a qualitative analysis of the outputs to address this issue and present our findings through cherry- and lemon-picked examples from our test set (Table~\ref{tab:cases}). In particular, we manually expected the outputs predicted by the best model configurations found in Table~\ref{tab:ablation_study} (layers number = 1, layer dimension = 512).

The first two rows of Table~\ref{tab:cases} are illustrative examples of categories of  intent~\textendash~snippet pairs that the model can successfully translate. 
The first row demonstrates the ability of the model to generate multi-line snippets from a relatively abstract intent. 
The example in the second row shows the model's ability to properly use the instruction \texttt{lea} with the correct addressing mode (specified by the bracket [] in NASM syntax) to translate the intent. We note here that a1though the output would be considered incorrect based on automated metrics (e.g.  BLEU-4), it is considered correct using manual inspection.

We also highlight problems with the models through illustrative examples of failure outputs (Rows 3 and 4, Table~\ref{tab:cases}).
In the third row of the table, the model generates the wrong instruction due to the model's failure in using implicit knowledge (i.e. the bit-wise inversion to negate the contents of the register) because it was not explicitly mentioned in the intent. 
Row 4 illustrates the model's failure in predicting the right command among fifteen different conditional jumps in the dataset (\texttt{jle} instead of \texttt{jge}) in an if-then statement.  
To summarize, the failures we observed are caused either by a lack of implicit intent knowledge, the model generating incorrect instruction/identifiers (i.e., register names, labels, etc), or even both.

\section{Ethical Considerations}
\label{sec:ethics}
Recognizing that attackers use exploit code as a weapon, it is important to specify that the goal of the \textit{proof-of-concept} (POC) exploits is not to cause harm but to surface security weaknesses within the software. Identifying such security issues allows companies to patch vulnerabilities and protect themselves against attacks. 

\emph{Offensive security} is a sub-field of security research that employs ethical hackers to probe a system for vulnerabilities or can be a technique used to disrupt an attacker. \emph{Automatic exploit generation} (AEG), an offensive security technique, is a developing area of research that aims to automate the exploit generation process and to explore and test critical vulnerabilities before they are discovered by attackers~\cite{aeg}. Indeed, studying exploits on compromised systems can provide valuable information about the technical skills, degree of experience, and intent of the attackers who developed or used them. Using this information, it is possible to implement measures to detect and prevent attacks \cite{arce2004shellcode}.

\section{Conclusion}
\label{sec:conclusion}
We address the problem of automated exploit generation through NLP. We use Neural Machine Translation to translate the natural language intents into assembly code.
The contribution in this work is a new dataset, \datasetname{}, containing $3,200$ pairs of instructions in assembly language code snippets and their corresponding intents in English. These assembly language snippets can be combined together to generate attacks or exploits on Linux OS running on Intel Architecture 32-bit machines.  

\datasetname{} represents a first step towards the ambitious goal of automatically generating shellcodes from natural language. Our experimental evaluation has shown promising early results, demonstrating the feasibility of generating assembly code instructions with high accuracy.

\section*{Acknowledgements}
This work has been partially supported by the University of Naples Federico II in the frame of the Programme F.R.A., project id OSTAGE. 

\bibliographystyle{acl_natbib}
\bibliography{anthology,bibliography}

\end{document}